\documentclass[aps,prl,reprint,superscriptaddress,showpacs,amsmath,floatfix]{revtex4-1}
\usepackage{braket,graphicx}
\usepackage[colorlinks=true,linkcolor=blue,citecolor=blue,urlcolor=blue]{hyperref}

\pdfoutput=1

\newcommand{\MPQ}{\affiliation{Laboratoire Mat\'eriaux et Ph\'enom\`enes Quantiques, Universit\'e Paris Diderot, Sorbonne Paris Cit\'e, CNRS-UMR 7162, Case
courrier 7021, 75205 Paris Cedex 13, France}}
\newcommand{\LPN}{\affiliation{Laboratoire de Photonique et Nanostructures, CNRS-UPR20, Route de Nozay, 91460 Marcoussis, France}}
\newcommand{\ROMA}{\altaffiliation[Present address: ]{Dipartimento di Fisica, Sapienza Universit\`a di Roma, Piazzale Aldo Moro, 5, I-00185 Roma, Italy}}

\begin{document}

\title{An electrically injected photon-pair source at room temperature}

\author{Fabien Boitier}\author{Adeline Orieux}\ROMA\author{Claire Autebert}\MPQ
\author{Aristide Lema{\^{i}}tre}\author{Elisabeth Galopin}\LPN
\author{Christophe Manquest} \author{Carlo Sirtori} \author{Ivan Favero} \author{Giuseppe Leo} \author{Sara Ducci}\email{sara.ducci@univ-paris-diderot.fr}\MPQ

\date{\today}

\pacs{42.65.Lm, 03.67.Bg, 42.55.Px, 42.82.-m}

\begin{abstract}
One of the main challenges for future quantum information technologies is miniaturization and integration of high performance components in a single chip. 
In this context, electrically driven sources of non-classical states of light have a clear advantage over optically driven ones. 
Here we demonstrate the first electrically driven semiconductor source of photon pairs working at room temperature and telecom wavelength. 
The device is based on type-II intracavity Spontaneous Parametric Down-Conversion in an AlGaAs laser diode and generates pairs at 1.57 $\mu$m. 
Time-correlation measurements of the emitted pairs give an internal generation efficiency of  $7\times 10^{-11}$ pairs/injected electron. 
The capability of our platform to support generation, manipulation and detection of photons opens the way to the demonstration of massively parallel systems for complex quantum operations.
\end{abstract}

\maketitle
Photons have a peculiar advantage in the development of quantum information technologies~\cite{ladd_quantum_2010,gisin2007,eisaman_invited_2011}, since they behave naturally as flying qubits presenting a high speed transmission over long distances and being almost immune to decoherence~\cite{ursin2007entanglement,stucki2009continuous}. 
The intrinsic scalability and reliability of integrated photonic circuits has recently given rise to a new generation of  devices for quantum communication, computation and metrology~\cite{obrien_photonic_2009}.
Nevertheless even if great progress have been made in the manipulation~\cite{matthews2009manipulation,PhysRevLett.111.213603} and detection~\cite{sprengers2011waveguide} of nonclassical state of light on chip, a complete integration of the light source in the photonic circuitry stays one of the main challenges on the way towards large scale applications; such devices would have a clear advantage over optically driven ones in terms of portability, energy consumption and integration. 
Semiconductor materials are ideal to achieve extremely compact and massively parallel devices: concerning photon-pair sources, the bi-exciton cascade of a quantum dot has been used to demonstrate an entangled-light-emitting diode at a wavelength of 890\,nm~\cite{salter_entangled-light-emitting_2010}.
However, even if the use of a single emitter guarantees a deterministic emission, these devices operate at cryogenic temperature, greatly limiting their potential for applications.

Optical parametric conversion offers an alternative approach. Despite its non-deterministic nature, this process is the most widely used to produce photon pairs for quantum information and communications protocols. 
Up to now, entangled photon pairs have been generated by optical pumping in passive semiconductor waveguides by exploiting four-wave mixing in Silicon~\cite{matsuda_monolithically_2012} or SPDC in Aluminium Gallium Arsenide (AlGaAs)~\cite{Orieux_Bell_2013, Valles_2013}.
Thanks to its direct band gap, the latter platform presents an evident interest for the electrical injection. 
In order to deal with the isotropic structure of this crystal, several solutions have been proposed to achieve nonlinear optical conversion in {AlGaAs} waveguides~\cite{fiore_phase_1998, yu_growth_2007, lanco_semiconductor_2006, horn_monolithic_2012, van_der_ziel_phase-matched_1974}; among these, modal phase matching, in which the phase velocity mismatch is compensated by multimode waveguide dispersion, is one of the most promising to monolithically integrate the laser source and the nonlinear medium into a single device~\cite{orieux_laser_2012, bijlani_semiconductor_2013}.
In this scheme, the interacting modes can either be confined by homogeneous claddings~\cite{de_rossi_third-order-mode_2004} or by photonic band gap~\cite{bijlani_bragg_2009}, this latter option avoiding aging problems via the reduction of the total aluminum content.

\begin{figure}
	\includegraphics[width=80mm]{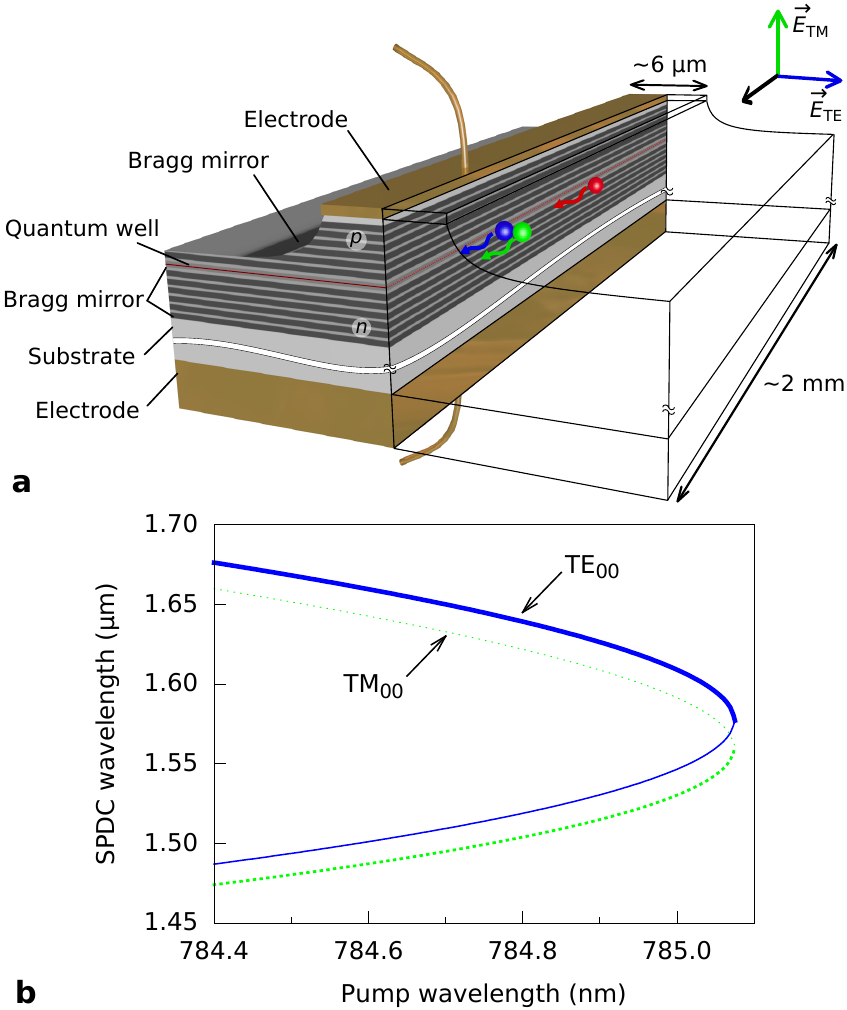}
	\caption{\label{1}(Color online) %
	Working principle of the device. (a)~Schematic view of the source. The laser light emitted by the quantum well is converted into telecom photon pairs by intracavity spontaneous parametric down-conversion. (b)~Simulated tuning curves of the type-II phase matching at $T=20^{\circ}$C. Energy conservation imposes pair generation either on the thick or the thin branches of the curves.}
\end{figure}

In this letter we present an electrically injected AlGaAs device that emits photons pairs at telecom wavelength and operates at room temperature.
Our device, shown in Fig.~\ref{1}(a), has been engineered for simultaneous lasing around 785\,nm and efficient type-II internal SPDC with photon pairs around 1.57\,$\mu$m.
Two Bragg mirrors provide both a photonic band gap vertical confinement for the laser mode --\,a Transverse Electric Bragg (TEB) mode\,-- and total internal reflection claddings for the photon-pairs modes (one TE$_{00}$ and one TM$_{00}$). 
The nonlinear process is possible thanks to the interaction of the TEB pump mode and the two twin photon modes verifying the equations of energy conservation and type-II phase matching:
\begin{eqnarray*}
	\hbar\omega_{\mathtt{\,TEB}} &=& \hbar\omega_{\mathtt{\,TE}_{00}}+\hbar \omega_{\mathtt{\,TM}_{00}}\\
	n_{\mathtt{\,TEB}}(\omega_{\mathtt{\,TEB}})\,\omega_{\mathtt{\,TEB}}
	&=& n_{\mathtt{\,TE}_{00}}(\omega_{\mathtt{\,TE}_{00}})\, \omega_{\mathtt{\,TE}_{00}} + n_{\mathtt{\,TM}_{00}}(\omega_{\mathtt{\,TM}_{00}})
	 \,\omega_{\mathtt{\,TM}_{00}}
\end{eqnarray*}
where $\omega_{i}$ and $n_{\mathrm{i}}$ (with $i=\mathrm{TEB},$ $\mathrm{TE}_{00},\mathrm{TM}_{00}$) are the angular optical frequency and the effective index of the $i$-th mode. 
The simulated tuning curves based on Ref.~\cite{chilwell_thin-films_1984,gehrsitz_refractive_2000}, solutions of the above system, are shown in Fig.~\ref{1}(b). 
Due to the strong dispersion of the TEB mode arising from the proximity to the energy band gap of the waveguide core, small shifts of the laser wavelength from degeneracy produce a large wavelength separation between the generated photons.
For this reason, taking into account the sensitivity range of our single-photon avalanche photodiodes, our spectral window to detect the two photons of each pair is limited to the region of frequency degeneracy. 

The sample was grown by molecular beam epitaxy on a (100) {\em n}-doped GaAs substrate. It consists of a {\em n}-doped 6-period Al$_{0.80}$Ga$_{0.20}$As/Al$_{0.25}$Ga$_{0.75}$As Bragg reflector (lower mirror), a 298\,nm Al$_{0.45}$Ga$_{0.55}$As core with a 8.5\,nm Al$_{0.11}$Ga$_{0.89}$As quantum well (QW) in the middle, and a {\em p}-doped 6-period Al$_{0.25}$Ga$_{0.75}$As/Al$_{0.80}$Ga$_{0.20}$As Bragg reflector (upper mirror).
The Bragg reflectors are gradually doped from $1 \times 10^{-17}$ cm$^{-3}$ to $2 \times 10^{-18}$ cm$^{-3}$. 
A 230\,nm GaAs cap layer ($2 \times 10^{-19}$ cm$^{-3}$ {\em p}-doped) protects the structure and facilitates the upper contact.
Waveguides are fabricated using wet chemical etching to define 5.5-6\,$\mu$m wide and 2\,$\mu$m deep ridges along the (011) crystalline axis, in order to exploit the maximum non-zero nonlinear coefficient and a natural cleavage plane. 
Processing is completed by sample thinning and contact metallization with Au alloys. Samples are cleaved into 2\,mm long stripes.

\begin{figure}
	\includegraphics[width=80mm]{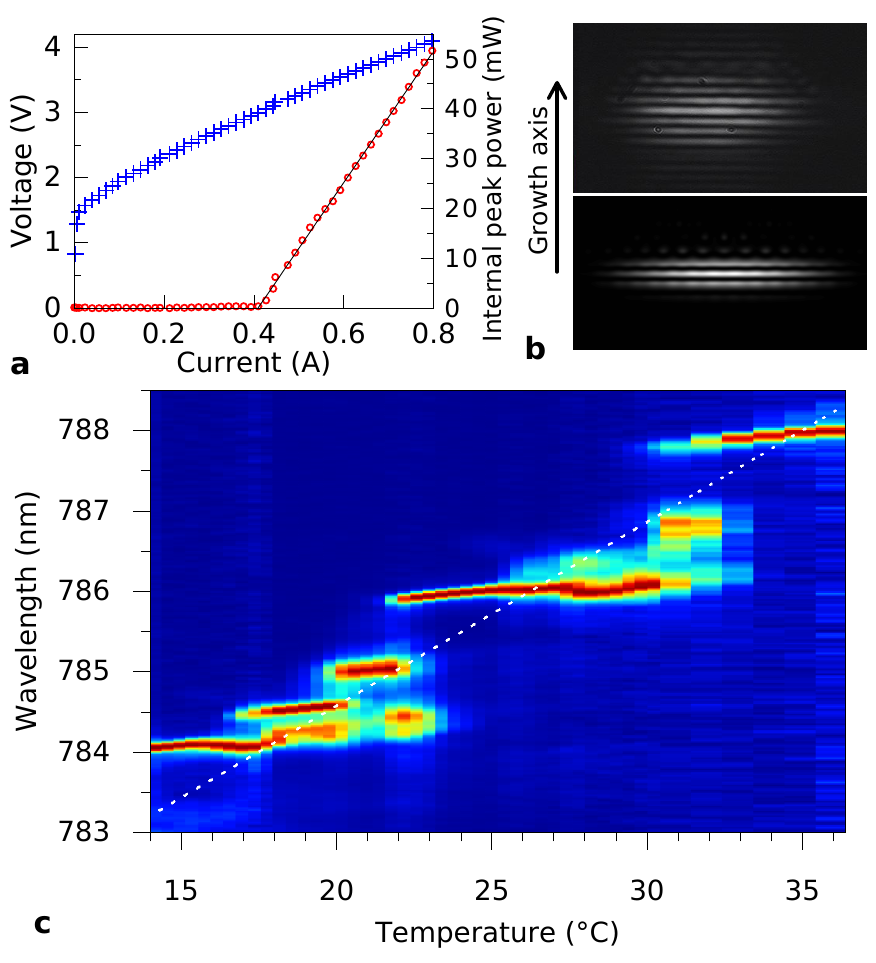}
	\caption{\label{2}(Color online) %
	Laser operation. %
	(a)~Voltage (cross) and internal optical power (circle) versus current. Measurements are performed with a current pulse duration of 120\,ns and a repetition rate of 10\,kHz for a heat-sink temperature of $19^\circ$C. The solid line is a linear fit for current values above laser threshold. The laser diode has an electrical resistivity of 3.1\,$\Omega$, a turn-on voltage of 1.6\,V, a laser threshold of 0.420\,A and an efficiency of 267\,mW/A. (b)~Measured (top) and simulated (bottom) near-field emission of the laser mode. (c)~Normalized laser emission intensity as a function of wavelength and heat-sink temperature measured with a fibred optical spectrum analyser. The dashed line shows the expected temperature variation of the QW bandgap. }
\end{figure}

Figure~\ref{2}(a) shows the internal peak power and voltage characteristics of the device as a function of the injected current. 
The device is mounted epi-side-up on a copper heat-sink; the temperature can be tuned between 15 and 40$^{\circ}$C with a standard Peltier module. 
In order to avoid unwanted thermal drifts, we employ current pulses of duration 120\,ns and the repetition rate is set to 10\,kHz. 
The laser internal peak power is evaluated by taking into account the modal reflectivity of the TEB mode (79$\%$), numerically simulated by 2D FDTD. 
We observe that the turn-on voltage is $\sim$1.6\,V, which is very close to the QW bandgap ($\sim$1.58\,eV), thus meaning that no current-blocking effects occur at the hetero-interfaces. 
The threshold current is around 420\,mA, corresponding to a threshold currentdensity of 3.3\,kA/cm$^{2}$. 
This value is higher than state-of-the-art laser diodes in this spectral range~\cite{coldren2012diode} probably because of the crudely optimized doping of the Bragg mirrors. 
The spatial intensity distribution of the laser beam is studied by imaging the output facet; the recorded near-field distribution is reported in Fig.~\ref{2}(b) together with the corresponding numerical simulation, showing a clear evidence of emission on the TEB mode. 
Figure~\ref{2}(c) displays the laser emission intensity spectra as a function of heat-sink temperature, for an injected current of 650\,mA. 
Apart from the longitudinal mode hopping --\,typical of laser diode\,--, the general trend corresponds to the theoretical temperature dependence of the QW bandgap (0.23\,nm/$^{\circ}$C).

\begin{figure}
	\includegraphics[width=80mm]{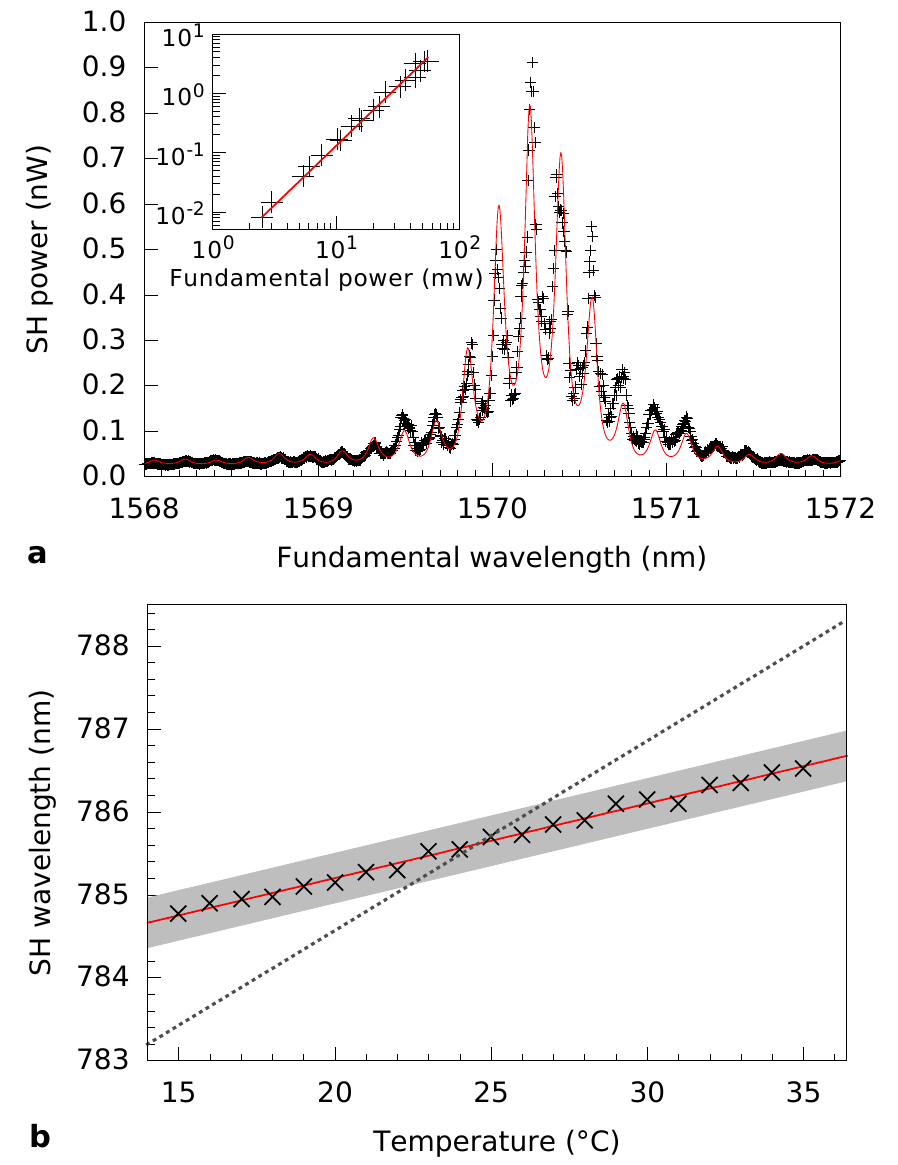}
	\caption{\label{3}(Color online) %
	Second harmonic generation. %
	(a) SH spectrum as a function of the fundamental wavelength at $T= 19 ^\circ$C. The curve is a fit taking into account propagation losses and modal facet reflectivities. The inset shows the peak SH power as a function of the fundamental beam power. The solid line shows the expected squared power law function. %
	(b) SH peak wavelength versus temperature. The solid line is a linear fit of the experimental data whereas the shaded area indicates the FWHM of the phase-matching bandwidth. The experimental slope of 0.09\,nm/K is consistent with the theoretical slope of our numerical modeling (0.07\,nm/K). The dashed line reports the expected variation of the QW bandgap presented in Fig.~\ref{2}(c).%
	}
\end{figure}

Optical propagation losses in the waveguide, a key issue for photon sources intended for quantum information, are measured via a standard Fabry-Perot technique~\cite{de_rossi_measuring_2005}: the values obtained for the TE$_{00}$ and TM$_{00}$ modes in the telecom range are around 2\,cm$^{-1}$. 
Similar measurements on an undoped waveguide giving a value of 0.1\,cm$^{-1}$, the losses on the active device are mainly attributed to doping. 
The nonlinear optical properties of the sample are first explored through a Second Harmonic (SH) generation measurement performed without electrical injection. 
An input beam at the fundamental wavelength is polarized at 45$^\circ$ and is injected in the waveguide in order to couple TE and TM modes simultaneously. 
Figure~\ref{3}(a) shows a clear growth of the SH power for an input beam wavelength around 1.57\,$\mu$m at $T= 19^\circ$C; the inset shows the expected quadratic dependence of the SH power with the fundamental power. 
The observed modulation as a function of the input wavelength is due to Fabry-Perot interferences between the waveguide facets. 
The solid curve results from a fit taking into account propagation losses and modal reflectivities of the three interacting modes~\cite{Sutherland1996}. 
The inferred internal SH generation efficiency is $\sim$~35\,\%W$^{-1}$cm$^{-2}$ and the FWHM of the phase-matching bandwidth is $\sim$~0.6\,nm. 
Figure~\ref{3}(b) reports the variation of the SH peak wavelength with temperature. 
The comparison between these data and those of Fig.~\ref{2}(c) shows that the tunability curves of the laser emission and of the SH signal intersect in the explored temperature range.

\begin{figure}
	\includegraphics[width=80mm]{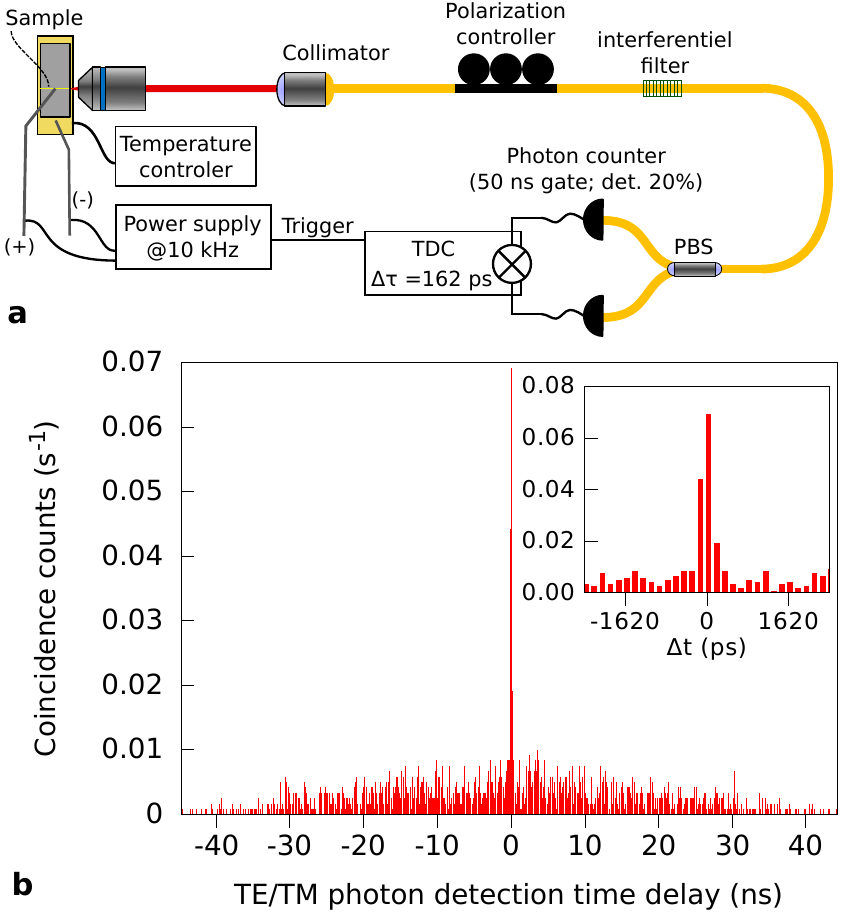}
	\caption{\label{4}(Color online) %
	Coincidence measurement.
	(a)~Experimental setup. 
	The emitted signal, collected through a 63X microscope objective, is focussed into a fibred 1.2\,nm-FWHM interferential filter centered at 1.57\,$\mu$m then sent in a fibred polarizing beam-splitter. 
	The emerging TE and TM photons are detected with two InGaAs single-photon avalanche photodiodes having 20\% detection efficiency and 50\,ns gate, synchronized with the current pulses. 
	A time-to-digital converter is used to analyze the time correlations between detected photons.
	(b)~Time-correlation histogram of TE/TM photons around 1.57\,$\mu$m at T$=25^{\circ}$C. 
        The sample is electrically injected with current pulses having an intensity of 700\,mA, a duration of 60\,ns and a repetition rate of 10\,kHz. 
        The data were accumulated during 1200\,s with a sampling resolution of 162\,ps. 
        The inset shows a zoom on the sharp central peak.%
	}
\end{figure}

In order to confirm the existence of a working region of the device and to demonstrate the emission of photon pairs around 1.57\,$\mu$m, time-correlation measurements are performed under electrical injection (see Fig.~\ref{4}(a)).
The detected SPDC signal is optimized by tuning the temperature. 
Figure~\ref{4}(b) shows a histogram of the detection time delays between TE and TM polarized photons at $T= 25^\circ$C. 
The sharp peak emerging from the background is a clear evidence of pairs production. 
From these data, taking into account the overall transmission along the optical path, we can estimate that the internal generation efficiency of the device is $\sim 7\times 10^{-11}$ pairs per injected electron above the threshold.
This value corresponds to a SPDC efficiency $\sim 10^{-9}$\,pairs/pump photon: 
these results are in agreement with our SH generation efficiency, letting expect $\sim 6\times 10^{-9}$\,pairs/pump photon, and consistent with our numerical simulation on an undoped structure giving $\sim 1.8\times 10^{-8}$\,pairs/pump photon for a 2\,mm-long waveguide. 
Note here that such efficiency compare well with those obtained in a completely passive device based on the same kind of phase matching~\cite{horn_monolithic_2012}.

The Signal to Noise Ratio (SNR) is evaluated by taking the number of true coincidences within the FWHM of the peak over the background signal on the same time window; data presented in Fig.~\ref{4}(b) give a SNR of 13.5, mainly limited by the luminescence noise of the device. 
In this respect, an optimization work leading to smaller laser threshold will be beneficial to reduce spurious luminescence and, thus, to increase the SNR. 
Our result enables to estimate the fidelity F to the Bell state $\ket{\psi_+}$ that can be produced with our device. 
Assuming that the source emits a Werner state~\cite{werner_quantum_1989,barbieri_generation_2004}--\,which is reasonable since the noise is not polarized\,--, the associated density matrix is $\hat{\rho}_W =P\ket{\psi_+}\bra{\psi_+} +(1-P)/4 \times \openone$ with $P= SNR/(2+SNR)$. 
This leads to a maximal fidelity estimation F to $\ket{\psi_+}= (1+3P)/4 \sim$ 90\,\%, which is compliant with future experimental violation of Bell's inequality.

These results open the way towards large scale photonic-circuit-based quantum computation. 
Indeed one application of this source could be the controlled on-chip electrical injection of an arbitrary number of heralded single photons or photon pairs on an arbitrary number of input modes of an integrated photonic circuit. 
This could be achieved by fabricating a monolithic device consisting of equally spaced laser diodes independently injected through a control electronics, which is allowed by the mature III-V technology.
Interfacing it with multiport reconfigurable circuits~\cite{ladd_quantum_2010} would allow practical medium size reconfigurable on-chip quantum photonics computation, such as boson sampling~\cite{crespi2013integrated,tillmann2013experimental,broome2013photonic,spring2013boson} and multiple photon quantum walks allowing medium size optical simulations~\cite{aspuru2012photonic}. 

\begin{acknowledgements}
This work was partly supported by R\'egion Ile-de-France in the framework of C\textquoteright Nano IdF with the TWILIGHT project.
F.B. acknowledges the Labex SEAM `Science and Engineering for Advanced Materials and devices' for financial support.
A.O. et C.A. acknowledge the D\'el\'egation G\'en\'erale de l'Armement (DGA) for financial support.
We acknowledge G. Boucher and A. Eckstein for help with the experimental setup and A. Andronico for discussions on numerical simulations.
S.D. and C.S. are members of Institut Universitaire de France.
\end{acknowledgements}

\end{document}